\documentclass[preprint,12pt,showkeys,nofootinbib,floatfix]{revtex4}
\usepackage{amssymb}
\usepackage{amsfonts}
\usepackage{amsmath}
\usepackage{graphicx}
\usepackage{subfigure}
\usepackage[dvipdfm,
            pdfstartview=FitH,
            bookmarksnumbered=true,
            bookmarksopen=true,
            colorlinks,
            pdfborder=001,
            linkcolor=green,
            anchorcolor=green,
            citecolor=red
            ]{hyperref}
\usepackage[toc,page,title,titletoc,header]{appendix}
\usepackage{graphics}
\usepackage{epsfig}
\usepackage{slashed}
\usepackage{latexsym}
\usepackage{syntonly}
\newcommand{\f}{\begin{equation}}
\newcommand{\ff}{\end{equation}}
\newcommand{\fa}{\begin{eqnarray}}
\newcommand{\ffa}{\end{eqnarray}}
\newcounter{fignr}

\begin{document}
\title{Thermal geometry from CFT at finite temperature}
\author{Wen-Cong Gan$^{1,2}$}
\thanks{E-mail address:ganwencong@gmail.com}
\author{Fu-Wen Shu$^{1,2}$}
\thanks{E-mail address:shufuwen@ncu.edu.cn}
\author{Meng-He Wu$^{1,2}$}
\thanks{E-mail address:menghewu.physik@gmail.com}
\affiliation{
$^{1}$Department of Physics, Nanchang University, Nanchang, 330031, China\\
$^{2}$Center for Relativistic Astrophysics and High Energy Physics, Nanchang University, Nanchang 330031, China}
\begin{abstract}
We present how the thermal geometry emerges from CFT at finite temperature by using the truncated entanglement renormalization network, the cMERA. For the case of $2d$ CFT, the reduced geometry is the BTZ black hole or the thermal AdS as expectation. In order to determine which spacetimes prefer to form, we propose a cMERA description of the Hawking-Page phase transition. Our proposal is in agreement with the picture of the recent proposed surface/state correspondence.
\end{abstract}
\maketitle
\section{Introduction}
Ever since AdS/CFT correspondence was proposed \cite{maldacena1}\cite{gkp}\cite{witten}, a great deal of attention has been paid to understand its basic mechanism. A recent progress was made through tensor network, which is a powerful tool to deal with system of multi-degree freedom, such as critical Ising model, whose continuum is a CFT. A particular significant tensor network is multi-scale entanglement renormalization ansatz (MERA)\cite{vidal1,vidal2}, containing disentanglers, which can remove short-range entanglement and thus produce an efficient description of critical system. Inspired by the Ryu-Takayanagi formula of the holographic entropy\cite{RT}, The relationship between MERA and AdS/CFT was first noticed by Swingle\cite{swingle}. He pointed out that the MERA that emerges from critical system (which can be viewed as discrete CFT) can be viewed as discrete time slice of AdS space, where the renormalization direction of MERA corresponds to the discrete radial direction of the AdS space. To apply MERA to field theory, we need to generalize MERA to continuous version. The generalization was made by Haegeman et al. in \cite{erqf} where they proposed the continuous MERA (cMERA) which can make entanglement renormalization for quantum fields in real space. Nozaki et al. applied it, for the first time, to the context of AdS/MERA, and the emergent holographic (smooth) geometry was obtained \cite{hg}.

\begin{figure}
  \centering
  \includegraphics[width=.8\textwidth]{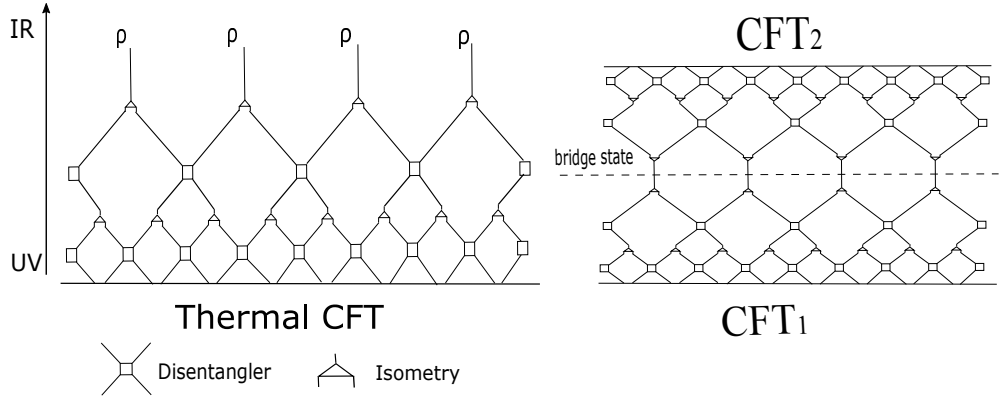}\hspace{4cm}
  \caption{The MERA applied to a thermal state truncates when it reaches maximally mixed state. (left) Double MERA network (right). At the center there is a bridge state which glues two copies of the standard MERA. This state is usually viewed as a black hole horizon.}\label{fig1}
\end{figure}

To discuss the correspondence between CFT and the spacetime containing black holes, it is necessary to consider CFT at finite temperature. In terms of the AdS/MERA, there are two different descriptions. One is based on the initial MERA and argues that after finite steps of entanglement renormalization, the MERA truncates at a level of multiple sites where the state at the top level (i.e. the truncated MERA state) becomes maximally mixed state and thus corresponds to the horizon of a black hole \cite{swingle}\cite{swingle2}. We often call it the truncated MERA. The alternative way is based on the thermofield double formalism\cite{maldacena2} and the emergent tensor network is often called double MERA. The double MERA composes of two copies of the MERA that connected by MPS state (Fig.\ref{fig1})\cite{mih}\cite{hm}\cite{jmv2}. A continuous extension of the double MERA has been achieved in \cite{hgf}, where the authors pointed out that it quantitatively agrees with a half of the AdS black hole spacetime. Even though the double MERA leads to some progress in discussing AdS/MERA at finite temperature, the truncated MERA still deserves more attention. This is based on the following two considerations: firstly, the continuous version of the truncated MERA, as a conjectured tensor network, whose validity to generate (smooth) black hole geometry has not yet been proved; secondly, the truncated MERA provides a possibility to give a cMERA description of the Hawking-Page phase transition\cite{hp}, which is lack of careful investigation in the past literatures.

In this letter, we would like to apply the truncated MERA to construct the holographic geometry of cMERA at finite temperature explicitly by studying the two dimensional CFT. Our results confirm that the thermal emergent geometry can be obtained in this way,  and point out that the emergent spacetime is BTZ black hole for sufficiently large temperature. On the other hand, we also try to give a cMERA description of the Hawking-Page phase transition. Our results show that depending on the cutoff parameters, it is possible to figure out the phase of the emergent spacetimes. The main result is consistent with the picture of the surface/state correspondence proposed recently in \cite{MT,mnstw}.

\section{Thermal spacetimes emerge from truncated cMERA}

Let us start with basic concepts of cMERA. The UV state in cMERA is defined as $|\Psi(u_{UV}) \rangle \equiv |\Psi \rangle$ and the IR state is defined as the state with no real space entanglement obtained by entanglement renormalization $|\Psi(u_{IR}) \rangle \equiv |\Omega \rangle$, where $u_{UV}=0$ and $u_{IR}=-\infty$. The states correspond to different $u$ are connected by unitary transformation:
\f
|\Psi(u) \rangle = U(u,u_{IR}) |\Omega \rangle,\ \ |\Psi(u) \rangle=U(0,u) |\Psi(u_{UV}) \rangle,
\ff
and
\f\label{ut}
U(u_1,u_2)=\mathcal P \exp \{-i\int_{u_2}^{u_1} (K(s)+L) \,\mathrm{d}s \},
\ff where $K(u)$ and $L$ are the continuous version of disentanglers and isometry, respectively. The isometry $L$ does not rely on $u$ because of the same course-graining procedures in every step of entanglement renormalization. The symbol $\mathcal{P}$ in \eqref{ut} is path-ordering symbol.

Now we would like to show that the thermal spacetimes, such as the BTZ black hole \cite{btz,bhtz} geometry can be emergent from a thermal CFT, with the help of the cMERA. As an example, let us consider a free scalar field in finite temperature. The thermal state in this system can be described by density matrix $\rho_0\equiv\rho(u=0)=e^{-\beta H}$, where
\f\label{H} H=\int d^{d}k \frac{1}{2}[\pi (\vec{k})\pi (-\vec{k})+(|\vec{k}|^2+m^2)\phi(\vec{k})\phi(-\vec{k})].\ff
The disentanglers and the isometry of this system is given by\cite{erqf}
\begin{eqnarray}
\label{K}K(s)&=&\frac{1}{2} \int d^{d}k \,g(k/\Lambda ,s)[\phi(\vec{k})\pi (-\vec{k})+\pi (-\vec{k}) \phi(\vec{k})],  \\
L&=&-\frac{1}{2} \int d^{d}x [\pi ({x})\vec{x}\cdot \vec\nabla_{{x}} \phi ({x})+\vec{x}\cdot \vec\nabla_{{x}} \phi ({x}) \pi ({x})\nonumber \\
&&+\frac{d}{2}
\phi ({x}) \pi({x}) +\frac{d}{2}\pi({x}) \phi ({x})],
\end{eqnarray}
where
\begin{eqnarray}
\label{g}g(k/\Lambda ,s)&=&\chi(s)\Gamma(|\vec k|/\Lambda), \\ \Gamma(x)&=&\Theta(1-|x|), \\\label{chi} \chi(s)&=&\frac{1}{2}\frac{e^{2s}}{e^{2s}+(\frac{m}{\Lambda})^2},
\end{eqnarray}
and $\Theta$ is the step function, and $\Lambda$ is a cutoff\cite{hg}.

In the context of the cMERA, $\rho_0$ is defined as the UV state of the network. Two nearby thermal states $\rho(u)$ and $\rho(u+du)$ which are connected by entanglement renormalization, are then related by
\begin{eqnarray}
\label{rho0}\rho_0 &=&U(0,u)\rho(u) U^{-1}(0,u) \\
\rho_0 &=&U(0,u+du)\rho(u+du) U^{-1}(0,u+du) \\
\label{rho}\rho(u) &=&U(u,u+du)\rho(u+du) U^{-1}(u,u+du).
\end{eqnarray}

As suggested by Swingle\cite{swingle}, the depth of the entanglement renormalization $u$ is associated with the extra dimension $z$ in bulk AdS spacetime.

For $d+1$ dimensional CFT, its gravity dual is $d+2$ dimensional AdS  $$ds_{AdS}^2= du^2+ \frac{e^{2u}}{\epsilon^2}(d\vec{x}^2-dt^2)=\frac{dz^2-dt^2+d\vec{x}^2}{z^2}$$
where $\epsilon$ is the UV cutoff, $z=\epsilon \cdot {e^u}$, $z$ is radial coordinate in Poincare patch and $u$ is identified with the renormalization direction of cMERA.

The MERAs for general QFT are expected to correspond to gravity dual with the metric
\begin{eqnarray}\label{a1}
ds^2= g_{uu} du^2+ \frac{e^{2u}}{\epsilon^2}d\vec{x}^2 +g_{tt} dt^2
\end{eqnarray}
The metric component for $\vec x$ is fixed because the coarse-graining procedure is the same. It was noticed in \cite{hg} that $g_{uu}$ is independent of $u$ if the QFT in question is CFT, otherwise, it should be a function of $u$.
To get the corresponding spacetime metric from the state of cMERA, it is helpful to introduce the quantum distance between states\cite{hg}
$$\mathcal{D}_{HS} (\rho_1,\rho_2)=\frac12 \mathrm{Tr} (\rho_1-\rho_2)^2, $$
where $\rho_1,\rho_2$ are the density matrices of two states. This is called Hilbert-Schmidt distance. In particular,  when the state is pure state, we have
$$\mathcal{D}_{HS}(\psi_1,\psi_2)=1-|\langle \psi_1|\psi_2\rangle|^2.$$

 For a set of quantum states parameterized by $\xi=(\xi_1,\xi_2 \cdots)$, the quantum distance between two nearby states $ \psi(\xi)$ and $\psi(\xi+d\xi)$ can be expressed as follows
$$\mathcal{D}_{HS}[\psi(\xi),\psi(\xi+d\xi)]=g_{ij}(\xi)d\xi_i d\xi_j,$$
where $g_{ij}$ is usually called the quantum metric (or the Fisher information metric).

In general, it is very difficult to produce the information metric from a QFT given, for instance, by \eqref{H}, due to the very complicated commutation relations of the operators. However, as far as a CFT (a massless free scalar field) is concerned, the situation is considerably simplified. From Eq. \eqref{chi} we have $\chi(s)=\frac12$, which implies that both $g$ and $K$ are independent of $u$ according to Eqs. \eqref{K} and \eqref{g}. From the definition of $U$ we see that $U$ is also independent of $u$. Combining Eqs. \eqref{rho0}-\eqref{rho}, this further indicates that $\rho_0$, $\rho(u)$ and $\rho(u+du)$ do not depend on $u$. The information metric $g_{uu} du^2= \mathcal{D}_{HS}(\rho(u),\rho(u+du))=\frac12 \mathrm{Tr} (\rho(u)-\rho(u+du))^2$ is therefore not a function of $u$ as well.

On the other hand, although it is quite complicated\footnote{The explicit expression is not necessary for our following discussions.}, we can expect that $g_{uu}$ should be a function of $\beta$, due to the fact that $\rho_0=e^{-\beta H}$. Without loss of generality, let us denote it by $f(\beta)$, i.e., $g_{uu}=f(\beta)$. After plugging it into (\ref{a1}), we get the spacetime geometry dual to a massless free scalar field at finite temperature
\begin{align}
ds^2= f(\beta) du^2+ \frac{e^{2u}}{\epsilon^2}d\vec{x}^2 +g_{tt} dt^2
\end{align}
Introducing a coordinate transformation, $z=\frac{1}{\Lambda} e^{-\sqrt{f}u}$, we get
\f\label{tads}
ds^2= \frac{dz^2}{z^2}+ \frac{1}{(\Lambda z)^{2/\sqrt{f}}}d\vec{x}^2 +g_{tt} dt^2,
\ff
which coincides the metric of pure AdS, except that it involves an extra parameter $\beta$. After making the following transformation\cite{FTE}
\fa
t-x&=& -e^{\theta-u}\sqrt{1-\frac{r_+^2}{r^2}}, \nonumber \\
t+x&=& e^{\theta+u}\sqrt{1-\frac{r_+^2}{r^2}}, \nonumber \\
z&=&\frac{r_+ e^{\theta}}{r},
\ffa
one can map it to the well-known BTZ black hole metric explicitely
\f\label{btz}
ds^2=R^2\left[-\left(\frac{r_+^2}{r^2}-1\right)du^2 + \frac{dr^2}{r^2-r_+^2} + \frac{r_+^2}{r^2}d\theta^2\right],
\ff
where $r_+$ denotes the horizon radius of the black hole.

To be honest, there is a subtlety in derivation of \eqref{btz}. Actually, the derived geometry could be a BTZ black hole as given by \eqref{btz} or a thermal AdS as given by \eqref{tads}. We will fix this subtlety in the next section.

\section{cMERA description of the Hawking-Page transition}

Even though the Hawking-Page phase transition has been intensively studied from the gravity side, it is still lack of full investigations from the point of view of MERA network and its continuous generalization. It was first pointed out in \cite{swingle} that the Hawking-Page phase transition could possibly be produced by MERA at finite temperature. The main idea is that depending on whether the system shrinks to a point or reaches infinite temperature (maximally mixed state) first under the entanglement renormalization, the emergent spacetime tends to be thermal AdS or AdS black hole respectively. One evidence of this argument was mentioned in \cite{swingle2} where the author considered a thermal state $\rho=e^{-\beta H}$ renormalized to a maximally mixed state $\rho'=e^{-\beta_0 H} (\beta_0 \approx 0)$ under continuous entanglement renormalization, calculated the total RG flow time from the UV state to the IR state, finding it in agreement with the distance from a UV cutoff $r=\delta$ to the horizon in the AdS black hole background.

\begin{figure}
  \centering
  \includegraphics[width=.6\textwidth]{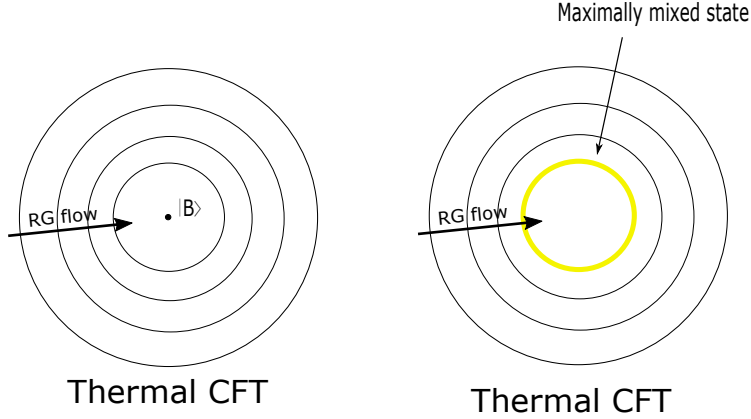}\hspace{3cm}
  \caption{Under the entanglement renormalization, if the system shrinks to a point (which corresponds to the boundary state $|B \rangle$) first, the emergent spacetime tends to be thermal AdS (left) or if the system reaches infinite temperature (maximally mixed state) first, and then it truncates at some level (the yellow circle in the picture), and the emergent spacetime tends to be AdS black hole (right). }\label{fig2}
\end{figure}

We generalize the above statement to cMERA description and improve the proposal by using a boundary state $|B\rangle$ which corresponds to a point in the bulk\cite{bs} and calculate the RG flow time from a thermal state $\rho=e^{-\beta H}$ in the UV to the boundary state in the IR. This generalization is consistent with the picture of the SS-duality, which states that the zero size closed surface (i.e. a point) is dual to a boundary state $|B\rangle$ while a closed topological non-trivial convex surface $\Sigma$ corresponds to a mixed quantum state $\rho(\Sigma)$. The picture of our proposal is sketched in Fig. \ref{fig2}.

For the free scalar field theory at finite temperature $T=\beta^{-1}$, it is described by
\f\label{rho1}
\rho_0 = e^{-\beta H}=\frac1{Z} \sum_k e^{-\beta E_k}|k\rangle \langle k|,
\ff
where $Z$ is the partition function and $|k\rangle=a_k^{\dagger}|0\rangle$.

On the other hand, as is pointed in \cite{bs}, the IR state has no real space entanglement, and can be represented by the boundary state $|B\rangle$ with a regularization, i.e. $|\Omega \rangle=e^{-\epsilon H} |B\rangle $, where $\epsilon$ is the UV cutoff (or lattice spacing). After tracing out the right-moving part, the reduced density matrix is
\begin{align}
\rho(u_{IR}) =\frac1{Z} \sum_k e^{-2\epsilon E_k}|k\rangle \langle k|.
\end{align}

For CFT, $K(u)$ is independent of $u$, implying $U(0,u_{IR})=e^{i(K+L)u_{IR}}$ defined in \eqref{ut}. Substituting this result into \eqref{rho0}, and adopting an assumption\cite{swingle2} that \emph{the entanglement renormalization procedure only changes the temperature parameter by $e^{u_{IR}}$}, we obtain
\f\label{rho2}
\rho_0=U(0,u_{IR}) \rho(u_{IR}) U^{\dagger}(0,u_{IR}) =\frac1 Z \sum_k e^{-2\epsilon E_k e^{u_{IR}}}|k\rangle \langle k|.
\ff
It is straightforward to show from \eqref{rho1} and \eqref{rho2} that
\f
e^{-\beta E_k} \sim e^{-2\epsilon E_k e^{u_{IR}}},
\ff
which implies $u_{IR}=\ln (\beta/2\epsilon)$.

Now we can discuss the possible phases of the emergent spacetimes by adopting the argument made by Swingle in \cite{swingle2}. For a CFT with given temperature $T=\beta^{-1}$ at UV, after the entanglement renormalization and finally stops in the IR with a maximally mixed state (black hole horizon), the renormalization parameter $u$ is given by \cite{swingle2}
\[
u=\ln (\beta/\beta_0),\ \beta_0 \approx 0.
\]
If $u_{IR}< u$, i.e. $2\epsilon > \beta_0$, then under the entanglement renormalization, the system reaches boundary state (which correspond to a point in the bulk) first, and the emergent spacetime of the dual theory is thermal AdS. Otherwise, if $u_{IR}> u$, i.e. $2\epsilon < \beta_0$, then the system reaches maximally mixed state (which corresponds to black hole horizon) first, and the emergent spacetime is AdS black hole (BTZ in $2+1$ dimensions).

\section{Conclusions and discussions}

In this paper, we have obtained the thermal emergent geometry by applying the truncated cMERA to $2d$ CFT at finite temperature. As expected, the reduced geometry is either the BTZ black hole or the thermal AdS, depending on the initial temperature of the CFT. In addition, inspired by the Swingle's argument on the dual Hawking-Page phase transition, we have proposed a cMERA description of the transition. In particular, our results show that according to the cutoff parameters $\epsilon$ and $\beta_0$, it is possible to distinguish the phases of the emergent spacetimes. The proposal is in agreement with the picture of the recent proposed surface/state correspondence\cite{mnstw}.

 Although we are focusing on the $(2+1)d$ emergent spacetimes, our formulation based on the cMERA can be applied to holographic construction of higher dimensional geometry, even for wide class of spacetimes which are not locally AdS. The formulation of the cMERA description of the Hawking-Page transition can be also generalized to more general case. As one interesting point, it is of particular interest to figure out the connections between cMERA and quark-gluon confinement-deconfinement phase transition as its close relationship with Hawking-Page phase transition\cite{witten}.

In order to offer more evidence for the mysterious connection between cMERA and AdS/CFT, there are a lot of work deserves deep investigation. For example, the explicit form of the metric component $g_{tt}$ is still missing in our proposal. It is instructive in \cite{hg} that if we employ cMERA on a Lorentz boosted time slice, then we can read off $g_{tt}$ in principle. Besides, how to naturally get metric component $g_{xx}$ is also of great importance. It was shown in \cite{mnstw} that this is possible to emerge the full metric on a time slice of AdS by employing the so-called surface/state correspondence \cite{MT}\cite{mnstw}. The generalization to the case with finite temperature is made in\cite{gsw}. However, how to generalize it to more general spacetimes is still unknown, especially when it is in the framework of modified gravity, such as the Ho\v{r}ava-Lifshitz (HL) gravity \cite{HL}. Actually, it was found that HL gravity is a minimal holographic dual for the field with Lifshitz scaling \cite{HL2}. Our recent works \cite{shu1,shu2} found that various Lifshitz spacetimes are possible even without matter fields. Therefore, in principle, extension of the cMERA to modified gravity is possible.

\begin{acknowledgments}
This work was supported in part by the National Natural Science Foundation of China under Grant No. 11465012, the Natural Science Foundation of Jiangxi Province under Grant No. 20142BAB202007 and the 555 talent project of Jiangxi Province.
\end{acknowledgments}


\begin{thebibliography}{99}
\bibitem{maldacena1} J. Maldacena, ``The large-N limit of superconformal field theories and supergravity", Adv. Theor. Math. Phys. {\bf2} (1998) 231 [Int. J. Theor. Phys. {\bf38} (1999) 1113] [hep-th/9711200].

\bibitem{gkp}S. S. Gubser, I. R. Klebanov and A. M. Polyakov, ``Gauge theory correlators from non-critical string theory," Phys. Lett. B {\bf428} (1998) 105 [hep-th/9802109].

\bibitem{witten}E. Witten, ``Anti-de Sitter space and holography," Adv. Theor. Math. Phys. {\bf2} (1998) 253 [hep-th/9802150].


\bibitem{vidal1}
G.Vidal,``Entanglement renormalization," Phys. Rev. Lett. {\bf99}, 220405 (2007)[cond-mat/0512165].

\bibitem{vidal2}
G.Vidal,  ``Class of quantum many-body states that can be efficiently simulated," Phys. Rev. Lett. {\bf101}, 110501 (2008)[arXiv:cond-mat/0605597].

\bibitem{RT}
S. Ryu, T. Takayanagi,  ``Holographic derivation of entanglement entropy from the anti-de sitter space/conformal field theory correspondence," Phys. Rev. Lett. {\bf96}, 181602 (2006) [arXiv:hep-th/0603001].

\bibitem{swingle}
B.Swingle, ``Entanglement renormalization and holography," Phys. Rev. D {\bf86}, 065007 (2012)[arXiv:0905.1317 [cond-mat]].

\bibitem{erqf}
J.Haegeman, T.J.Osborne, H.Verschelde and F.Verstraete, ``Entanglement renormalization for quantum fields in real space," Phys. Rev. Lett.
{\bf110}, 100402 (2013)[arXiv:1102.5524 [hep-th]].

\bibitem{hg}
M. Nozaki, S. Ryu, T. Takayanagi, ``Holographic geometry of entanglement renormalization in quantum field theories," JHEP {\bf1210} (2012) 193 [arXiv:1208.3469 [hep-th]]

\bibitem{swingle2}
B.Swingle, ``Constructing holographic spacetimes using entanglement renormalization," [arXiv:1209.3304 [hep-th]].

\bibitem{maldacena2}
J.Maldacena, ``Eternal black holes in anti-de Sitter," JHEP {\bf0304} (2003) 021 [hep-th/0106112].

\bibitem{mih} H. Matsueda, M. Ishihara and Y. Hashizume, ``Tensor Network and Black Hole," Phys. Rev.
D {\bf87}, (2013) 066002, [arXiv:1208.0206 [hep-th]].

\bibitem{hm}
T. Hartman and J. Maldacena, ``Time Evolution of Entanglement Entropy from Black Hole
Interiors," JHEP {\bf1305} (2013) 014 [arXiv:1303.1080 [hep-th]].

\bibitem{jmv2} J. Molina-Vilaplana and J. Prior, ``Entanglement, Tensor Networks and Black Hole
Horizons," Gen. Rel. Grav {\bf46} (2014) 1823, [arXiv:1403.5395 [hep-th]]

\bibitem{hgf}
A. Mollabashi, M. Nozaki, S. Ryu, and T. Takayanagi, ``Holographic Geometry of cMERA for Quantum Quenches and Finite Temperature," JHEP {\bf1403} (2014) 098 [arXiv:1311.6095 [hep-th]].

\bibitem{hp}
S. W. Hawking and D. N. Page, ``Thermodynamics Of Black Holes In Anti-De Sitter Space,"
Commun. Math. Phys. {\bf87}, 577 (1983).

\bibitem{MT}
M. Miyaji, T. Takayanagi, ``Surface/State Correspondence as a Generalized Holography," Prog. Theor. Exp. Phys. {\bf2015}, 073B03 [arXiv:1503.03542 [hep-th]].

\bibitem{mnstw}
M. Miyaji, T. Numasawa, N. Shiba, et al, ``Continuous Multiscale Entanglement Renormalization Ansatz as Holographic Surface-State Correspondence," Phys. Rev. Lett. {\bf115}, (2015) 171602 [arXiv:1506.01353 [hep-th]].

\bibitem{btz}
M.Banados, C.Teitelboim, J.Zanelli, ``Black hole in three-dimensional spacetime". Phys. Rev. Lett. {\bf69}, (1992) 1849 [hep-th/9204099].

\bibitem{bhtz}
M.Banados, M.Henneaux, C.Teitelboim, et al. ``Geometry of the 2+1 black hole," Phys. Rev. D {\bf48}, (1993) 1506 [gr-qc/9302012].

\bibitem{FTE}
M. Fujita, T. Takayanagi, E. Tonni, ``Aspects of AdS/BCFT," JHEP {\bf1111} (2011) 043 [arXiv:1108.5152 [hep-th]].

\bibitem{bs}
M.Miyaji, S.Ryu, T.Takayanagi and X.Wen, ``Boundary states as holographic duals of trivial spacetimes," JHEP {\bf1505} (2015) 152 [arXiv:1412.6226 [hep-th]].

%



\bibitem{gsw}
Wen-Cong Gan,Fu-Wen Shu and Meng-He Wu, surface/state construction for asymptotic AdS spacetime. work in progress.

\bibitem{HL} P. Ho\v{r}ava, ``Quantum Gravity at a Lifshitz Point," Phys. Rev. D {\bf79} (2009) 084008, [arXiv:0901.3775 [hep-th]].

\bibitem{HL2} T. Griffin, P. Ho\v{r}ava, and C. M. Melby-Thompson, ``Lifshitz Gravity for Lifshitz Holography," Phys. Rev. Lett. {\bf110} (2013) 081602 [arXiv:1211.4872 [hep-th]].

\bibitem{shu1} F.-W. Shu, K. Lin, A. Wang, and Q. Wu,  ``Lifshitz spacetimes, solitons, and generalized BTZ black holes in quantum gravity at a Lifshitz point," JHEP {\bf1404} (2014) 056 [arXiv:1403.0946 [hep-th]].

\bibitem{shu2} K. Lin, F.-W. Shu, A. Wang, and Q. Wu, ``High-dimensional Lifshitz-type spacetimes, universal horizons and black holes in Ho\v{r}ava-Lifshitz gravity," Phys. Rev. D {\bf91} (2015) 044003 [arXiv: 1404.3413 [hep-th]].

\end{thebibliography}
\end{document}